\documentclass[journal=jacsat,manuscript=article]{achemso}
\setkeys{acs}{articletitle = true}
\usepackage[version=3]{mhchem} 
\usepackage{chemformula}

\usepackage[version=3]{mhchem}

\author{Sebastian Klemenz}
\affiliation
{Department of Chemistry, Princeton University, Princeton, NJ, 08544, USA}
\author{Aurland K. Hay}
\affiliation
{Department of Chemistry, Wellesley College, Wellesley, MA, 02481, USA}
\author{Samuel M. L. Teicher}
\affiliation
{Materials Department and Materials Research Laboratory, University of California, Santa Barbara, California 93106, USA}
\author{Andreas Topp}
\affiliation
{Max-Planck-Institut
f\"ur Festk\"orperforschung, Stuttgart, 70569, Germany}
\author{Jennifer Cano}
\affiliation
{Department of Physics and Astronomy, Stony Brook University, Stony Brook, New York 11974, USA}
\alsoaffiliation
{Center for Computational Quantum Physics, The Flatiron Institute, New York, New York 10010, USA}
\author{Leslie M. Schoop}
\affiliation
{Department of Chemistry, Princeton University, Princeton, NJ, 08544, USA}
\email{lschoop@princeton.edu}
\phone{+1 609-258-9390}

\title[An \textsf{achemso} demo]
  {The role of delocalized chemical bonding in square-net-based topological semimetals}

\abbreviations{XRD, PPMS, TSM, DFT}

\begin{document}

\begin{abstract}

Principles that predict reactions or properties of materials define the discipline of chemistry. In this work we derive chemical rules, based on atomic distances and chemical bond character, which predict topological materials in compounds that feature the structural motif of a square-net. Using these rules we identify over 300 potential new topological materials. We show that simple chemical heuristics can be a powerful tool to characterize topological matter. In contrast to previous database-driven materials categorization our approach allows us to identify candidates that are alloys, solid-solutions, or compounds with statistical vacancies. While previous material searches relied on density functional theory, our approach is not limited by this method and could also be used to discover magnetic and statistically-disordered topological semimetals. 

\end{abstract}

\section{Introduction}

The most fundamental goal of modern solid-state chemistry is to link crystal structure to properties. In terms of electronic properties, we commonly distinguish between metals, semiconductors, and insulators (Figure \ref{intro}(a)). The emergence of ``topological matter'' as a field has added new members to the family of electronically distinct states of matter. One example is the topological semimetal (TSM), which is of interest because of its unusual electronic and optical properties. TSMs are known to display ultra-high electronic mobility and have been suggested as base materials for ultra-fast optical switches.\cite{schoop2018chemical,liang2015ultrahigh,chan2017photocurrents,weber2018directly,zhu2017robust,wang2017ultrafast} Therefore, currently, there is a strong interest in developing design principles, which chemists are uniquely equipped to construct. \cite{khoury2019new,seibel2015gold} 
To develop topological materials, chemical heuristics can be immensely useful.  For example, rules based on chemical composition were recently constructed to predict new topological materials in \textit{ABX} honeycomb compounds and to distinguish between topological insulators and topological semimetals.\cite{zhang2018topological}
Other chemical approaches have focused on electrides as candidates for topological matter\cite{hirayama2018electrides} as well as crystallographic approaches for tetrahedral transition metal chalcogenides\cite{zhou2017tetrahedral} and the use of chemical intuition for property-driven phase predictions.\cite{chemint, gui2019new,malyi2019realization}
Here we derive rules based on chemical bonding, crystal structure, and electron count to predict TSMs, which will also allow us to predict different topological phases.
 
\section{Electron Counting and Electronic Structure}
 
First, let us remind ourselves about some established chemical guidelines. For example, we intuitively know that diamond is an insulator, while graphite conducts electricity, despite its identical chemical composition. But what is the chemical heuristic that lets us guess correctly?
The answer lies in the number of valence electrons, their distribution, and their location. Main group elements usually follow the (8-N) electron rule (octet rule). Compounds following this rule are called electron-precise and have a filled valence shell. All their valence electrons are either located in two-electron-two-center bonds (2\textit{e}2\textit{c} bonds) or in non-bonding electron lone-pairs. In a solid, this localization often leads to a band gap. Metals, by contrast, are open-shell systems where the electrons are ``free'', resulting in metallic conductivity. A class of compounds that falls in between these two cases is Zintl phases. In 1929,\cite{zintl1929salzartige} Zintl noticed that these compounds are peculiar; they are semiconducting despite being synthesized from metals. Zintl phases, as defined by the classical Zintl-Klemm concept, are intermetallic phases containing polyanions or polycations which exhibit chemical bonds of covalent character.\cite{kauzlarich1996Zintl,nesper2014zintl} 
Electronically, Zintl phases can be understood in the context of a Peierls distortion. Simliar to Roald Hoffmann's bonding model of a 1D chain of hydrogen atoms,\cite{hoffmann1987chemistry} 2\textit{e}2\textit{c} bonds are formed in Zintl phases to stabilize the phase electronically, resulting in the appearance of a band gap.

\section{Semimetals and Half-Filled Bands}
\begin{figure}
  \includegraphics[width=0.80\textwidth]{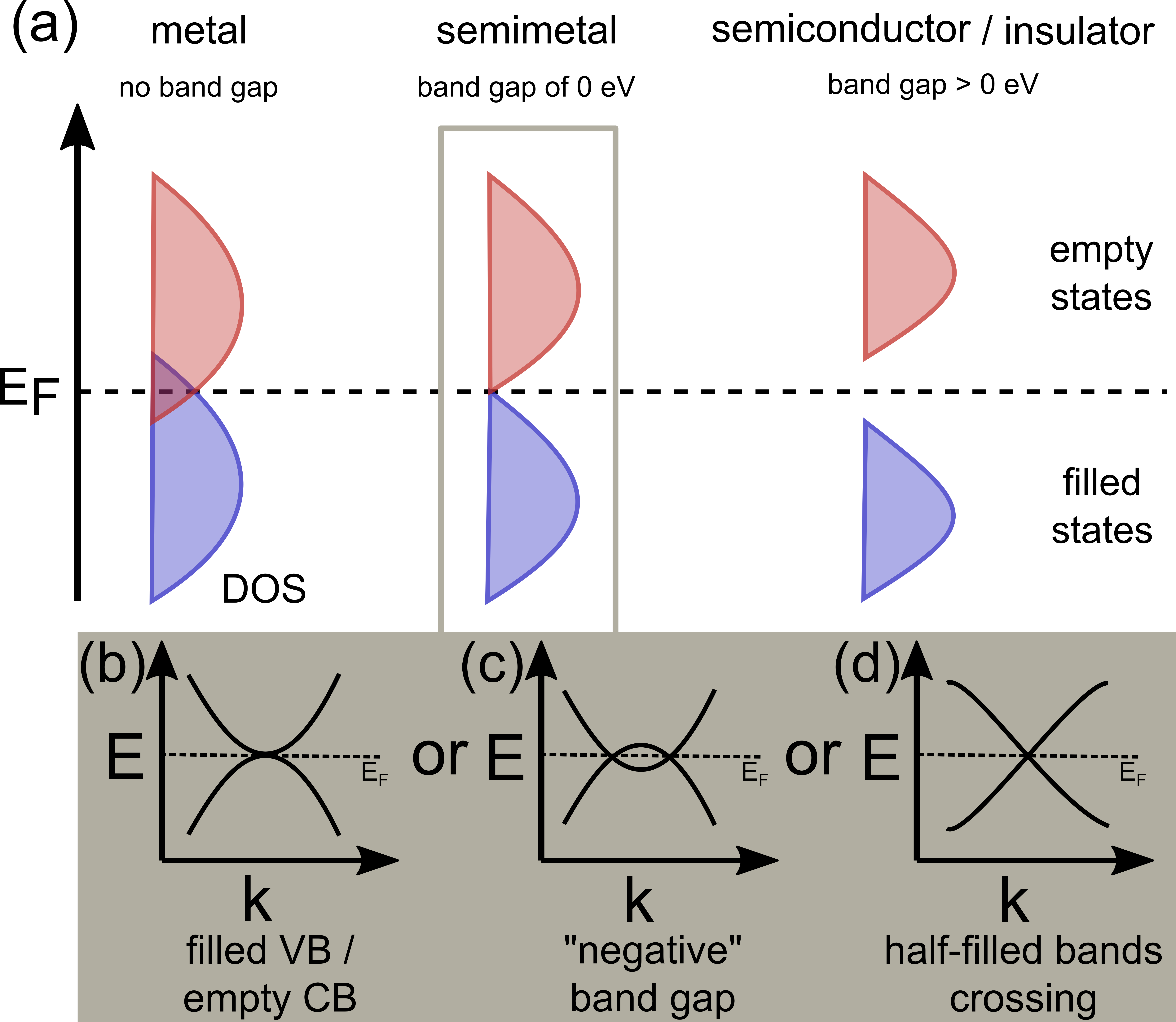}
  \caption{Schematic representation (a) of density of states (DOS) for material classes with different electron conduction behavior (metal, semimetal, and semiconductor/insulator). The added inlay (gray) shows different stylized band structure features leading to semimetallic behavior: Touching filled valence band (VB) and empty conduction band (CB) (b), overlapping bands (c) and crossing of half-filled bands (d).}
  \label{intro}
\end{figure}
 
\begin{figure}
  \includegraphics[width=0.90\textwidth]{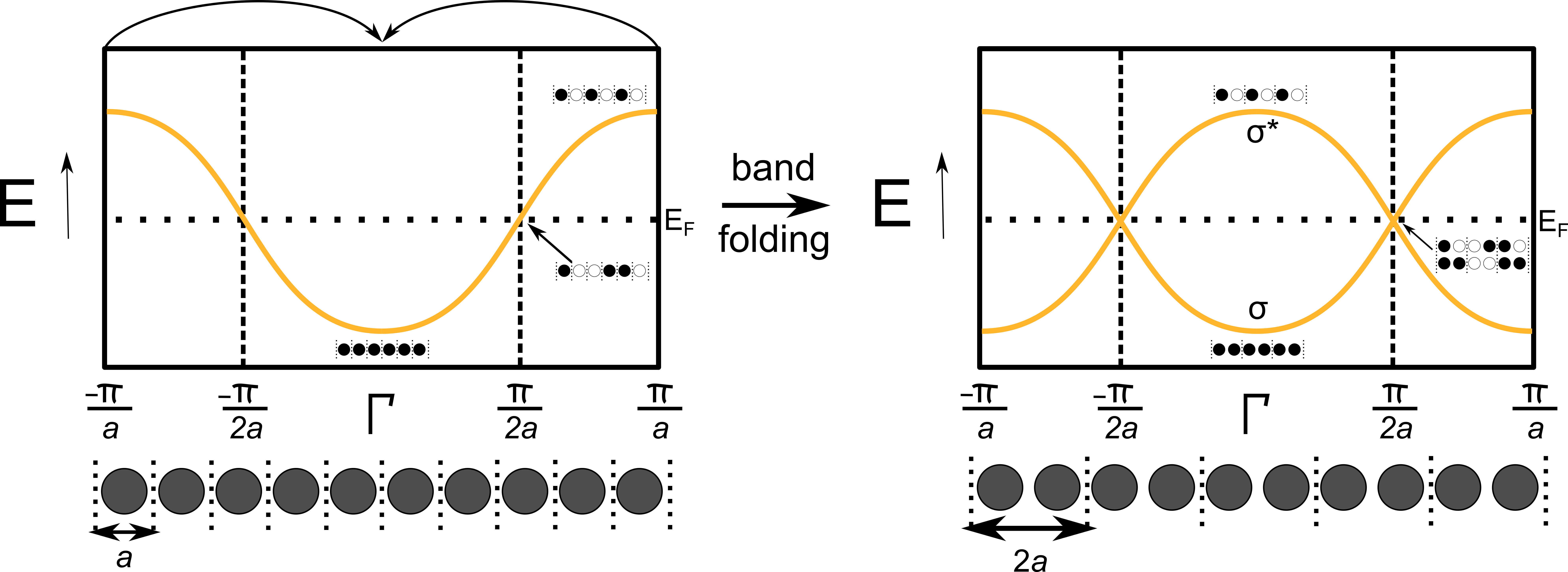}
  \caption{Band structures for a chain of s-orbitals before (left) and after (right) increasing the lattice parameter from a to 2a and the resultant band folding.}
  \label{bending}
\end{figure}

Semimetals are a class of materials in between metals and semiconductors (\mbox{Figure \ref{intro}}). They have no band gap, but the density of states (DOS) vanishes at the Fermi level (E$_\mathrm{F}$). Semimetals can be electron-precise like semiconductors when the filled and empty shells overlap or touch (Figure \ref{intro}(b,c)). But semimetals can also appear in materials where the valence band is half-filled (Figure \ref{intro}(d)). Intuitively, we would think that materials with half-filled bands are metals, so how can they be semimetals? We can understand this by considering, again, the model of the 1D chain of H atoms, where the DOS has a minimum at half filling (Figure \ref{bending}). The minimum at the Fermi level might be more obvious to see if the folded, but not yet Peierls distorted, band structure is considered; here the Fermi level only cuts through a single node of bands, resulting in a point-shaped Fermi surface. As we will explain later, the band folding can be related to a topological band structure.
 
When discussing systems with half-filled bands, the question of their stability might arise. Isolated half-filled bands are often unstable --- their electrons will either localize by undergoing a Peierls distortion (or charge density wave if more than one dimension is considered), or by forming a Mott insulator. There are, however, examples of stable half-filled bands; most famously graphene has a half-filled $p_z$-band. It is possible to chemically stabilize a half-filled band by delocalizing electrons (such as in aromatic systems or graphene).
In addition to aromatic systems, delocalized electrons can occur in chemical bonds that are different from 2\textit{e}2\textit{c} bonds. Such multi-center bonds can appear in molecules, cluster compounds, and in inorganic solids. An example for such multi-center bonds can be found in the I$_3^-$ anion, which has four electrons delocalized over three atoms (4\textit{e}3\textit{c} bond).\cite{landrum1997bonding} In solids, these bonds can be extended in one, two or three dimensions. This electronic delocalization is different from a ``free electron'' in metals; here a direction is associated with the delocalization, a consequence of the multi-center bond. Such ``hypervalent'' bonds have been described to exist in Sb chains and square-nets by Papoian and Hoffmann.\cite{a2000hypervalent} In their seminal paper, they extended the Zintl-Klemm concept by including hypervalent bonds, which removed the restriction of the octet rule and subsequently the necessity of a band gap in Zintl phases. The driving force for forming these hypervalent bonds is to obtain some degree of charge balance. This will result in a minimum in the density of states. Interestingly, extended hypervalent bonds require an electron count of one electron per bond meaning that the bond (or the band if we consider the solid) is half-filled. Therefore, these compounds should be semimetals if the band structure is folded.

\section{Hypervalency and Band Inversion}

We will argue here that the presence of an extended hypervalent bond in a solid can enforce a semimetallic band structure and therefore, hypervalent bonds can lead to electronic structures resembling TSMs. In some structural motifs, the hypervalent bond is accompanied by band folding. 

Such band foldings appear if two chemically equivalent atoms are in one unit cell (as in the folded band structure of a 1D chain of H atoms).
Band folding is important, since it causes the bands to cross. After the crossing point the bands will be ``inverted'', which is the key requirement for a band structure to be topological. 
We can understand a ``band inversion'' by considering graphene\cite{graphiteband}, the simplest example of a TSM. 
Graphene has a ``doubled'' unit cell containing two carbon atoms resulting in a folded, semimetallic, and inverted band structure.\cite{schoop2018chemical,graphiteband} The band structure consists of only two bands in the vicinity of the Fermi level, a $\pi$- and $\pi^\ast$-band that cross at a single point in the Brillouin zone (BZ) (the K point). The bands are inverted beyond the crossing point, meaning that the natural order of the bands ($\pi$ below $\pi^\ast$) is switched. 

In graphene, the energy region in which the inversion takes place is very large; the $\pi$- and $\pi^\ast$-bands are dispersed over tens of electron volts.
The band structure can be understood by considering the electrons in the half-filled $p_z$ band, which are delocalized over the whole 2D sheet. 

Thus, the band inversion is driven by chemical bonding, which is chemically different from a band inversion that results from overlapping filled and empty shells. The former type of band inversion usually extends over a larger range of energy, making those compounds superior topological materials. Such materials are easier to make (less doping and Fermi level tuning) and their properties are easier to measure making them more relevant for applications.
It is important to note that a band inversion is the defining aspect for \textit{all} topological phases, which includes topological insulators (TIs). Strong, weak, and higher-order TI phases result when spin-orbit coupling causes a band gap to open between two inverted bands.\cite{TQC,konig2007quantum,fu2007topological} The chemical rules introduced here for TSMs can be extended to other topological phases. 

\begin{figure}
  \includegraphics[width=0.80\textwidth]{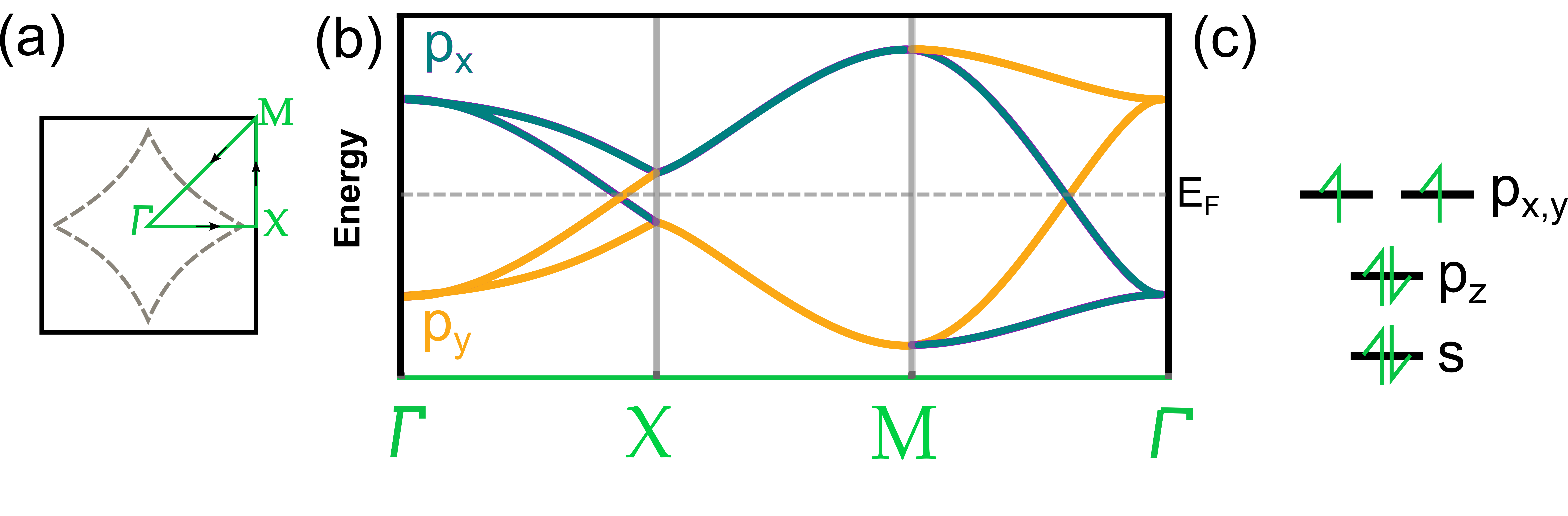}
  \caption{Brillouin zone with k-path (a) for the tight-binding model of bands originating from \textit{p$_x$}- and \textit{p$_y$}-orbitals on a $4^4$-net (b). The dashed line in (a) indicates the nodal-line. The dashed line in (b) indicates the filling of six electrons occupying the orbitals as shown in (c).}
  \label{ZrSiS}
\end{figure}

As argued above, extended hypervalent bonds fulfill the basic requirements for stabilizing inverted half-filled bands. One class of compounds that can feature hypervalent bonds are ``square-net materials''. These compounds feature the structural motif of a square-net. This square-net is side centered, i.e it contains two atoms per unit cell like graphene and is often referred to as the 4$^4$-net in crystallography literature. \cite{tremelhoffmannsquare,nuss2006geometric,charkin2007crystallographic}
Papoian and Hoffmann derived that such 4$^4$-nets can be hypervalently bonded with an electron count of six electrons per net atom.\cite{a2000hypervalent} This results in a half-filled band - the \textit{s}- and \textit{p$_z$}-bands will be filled, while the \textit{p$_x$}/\textit{p$_y$}-bands, which have to be degenerate at certain high symmetry points, are half-filled.\cite{Jen_Paper} In combination with the fact that a $4^4$-net contains two atoms per unit cell, this fulfills all our chemical requirements for a TSM. Figure \ref{ZrSiS} shows a simple tight-binding band structure derived for \textit{p$_x$}- and \textit{p$_y$}-orbitals on a site centered square-net, where both nearest neighbor and next-nearest neighbor interactions are considered (- details of the model can be found in ref.\,\cite{Jen_Paper}). At half-filling, the band structure resembles that of a TSM: the \textit{p$_x$}- and \textit{p$_y$}-bands cross and are inverted after the crossing point. There are multiple band crossings in the BZ, which results in a line of crossing points rather than isolated points (see Figure \ref{ZrSiS}(a)). Such materials are called nodal-line semimetals.

Indeed, numerous square-net phases match the requirements for hypervalent bonds and have been shown to be topological semimetals. The most famous example is ZrSiS - a topological nodal-line semimetal with a strikingly large band inversion.\cite{ZrSiSschoop2016} 
Here, silicon forms the $4^4$-net and has six electrons, in conformity with the rules derived by Papoian and Hoffmann. The bands at the Fermi level have been shown to be derived from Si $p_x$- and $p_y$-orbitals hybridized with Zr $d$-orbitals stabilizing the structure (see Figure \ref{MXZ-MG_data}(c)).\cite{Square_Review} The band structure of ZrSiS around the Fermi level matches the tight-binding model remarkably well.

We now proceed to use the derived link between hypervalency and band inversions to find new TSMs. To identify more square-net materials with a hypervalent bond we could search for $4^4$-nets with six electrons per net atom. However, electron counting by itself can be ambiguous and demands additional experimental proof to verify assigned oxidation states. In addition, as shown by Papoian and Hoffmann\cite{a2000hypervalent}, the hypervalent bond can be stable for an electron count that deviates from six, although, for strong deviations the square-net will distort.\cite{lei2019charge,a2000hypervalent,patschke2002polytelluride} Another approach is to consider atomic distances.\cite{hulliger1968new} If the atoms within the $4^4$-square-net are chemically bonded to each other, the interatomic distance within the net must be reduced compared to fully ionic square-nets. We previously showed that geometrical constraints can be applied to filter TSMs from electron-precise materials in PbFCl-type phases by extracting the inter-atomic distances of the $4^4$-net atoms and comparing them to the distance of the square-net to the next layer.\cite{Square_Review} 
Here, we link these geometric constraints to extended hypervalent bonds and additionally expand the ``tolerance factor'' to other square-net-based structure types. We show that the tolerance factor is a powerful and simple tool to predict new TSMs. The newly discovered TSMs reported here include off-stoichiometric and magnetic compounds that are not included in recent data-mining approaches because they go beyond the standard generalized gradient approximation (GGA)-based density functional theory (DFT). \cite{materiae,vergniory2019TMD} Therefore, this work emphasizes the strength of chemical heuristics for predicting topological phases of matter.

\section{Crystallographic Identification of Square-Net-based TSMs}

\begin{figure}
  \includegraphics[width=0.80\textwidth]{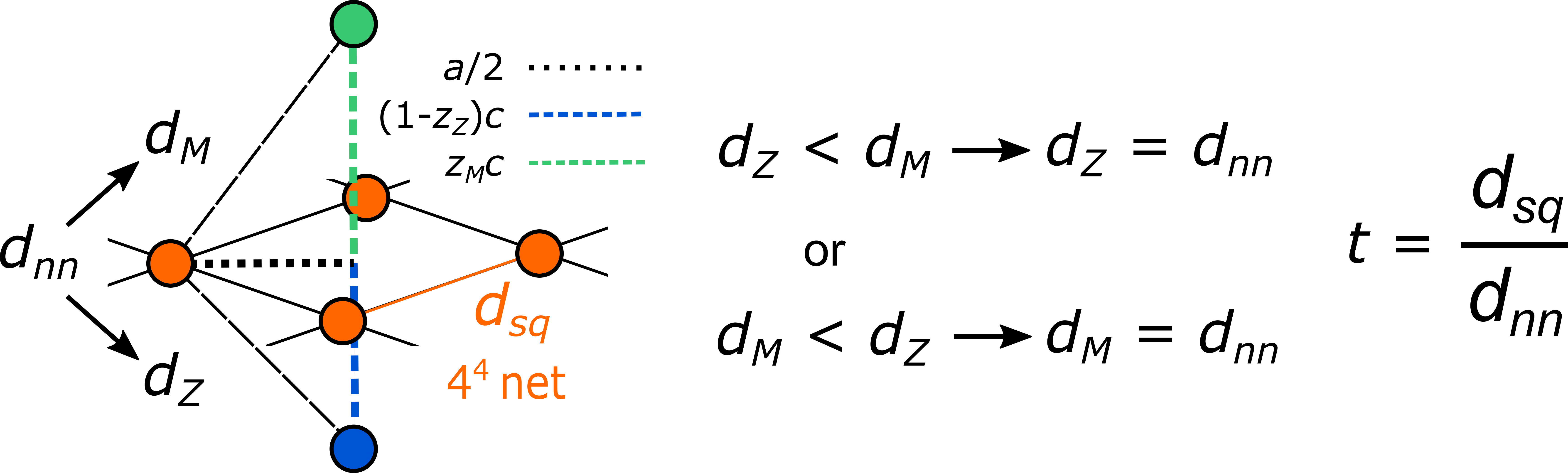}
  \caption{Graphic description of the tolerance factor $t = d_\mathrm{sq}/d_\mathrm{nn}$ and the definition of $d_\mathrm{nn}$ as the smaller of the two distances $d_\mathrm{M}$ and $d_\mathrm{Z}$.}
  \label{t-explained}
\end{figure}

The tolerance factor (\textit{t}) is defined as the ratio of the distance of atoms in the 4\textsuperscript{4}-net ($d_\mathrm{sq}$) and the distance to the nearest neighboring atom of a different layer in the structure ($d_\mathrm{nn}$) as shown in Figure \ref{t-explained}.
This is based on the assumption that the bands at the Fermi level are composed of \textit{p}-orbitals, conforming with the electron count of six electrons per square-net atom and resulting in a filled \textit{s}- and \textit{p$_z$}-band as well as a half-filled \textit{p$_x$}- and \textit{p$_y$}-band. Thus, it is only relevant for compounds where the $4^4$-net is composed of a main group (MG) element. Nonetheless, we found that compounds with a transition metal (TM) on the $4^4$-net position are mostly filtered out by the tolerance factor. Formula-agnostic tolerance factor screening can therefore be applied to all square-net materials in a given structural database.
 
In general, the band structure can be well described by the simple tight-binding model (TB model) of a square-net if the $4^4$-net is rather isolated. Isolation increases with decreasing \textit{t}, where in the extreme case of \textit{t} = 0 the square-net would be fully isolated. Thus, the smaller \textit{t} can be in real materials, the more ideal TSM band structures can be expected. We chose a cut off of \textit{t} $\leq$ 0.95 to separate materials that we expect to have an inverted band structure from materials that should have a non-inverted band structure.

\begin{figure}
  \includegraphics[width=0.80\textwidth]{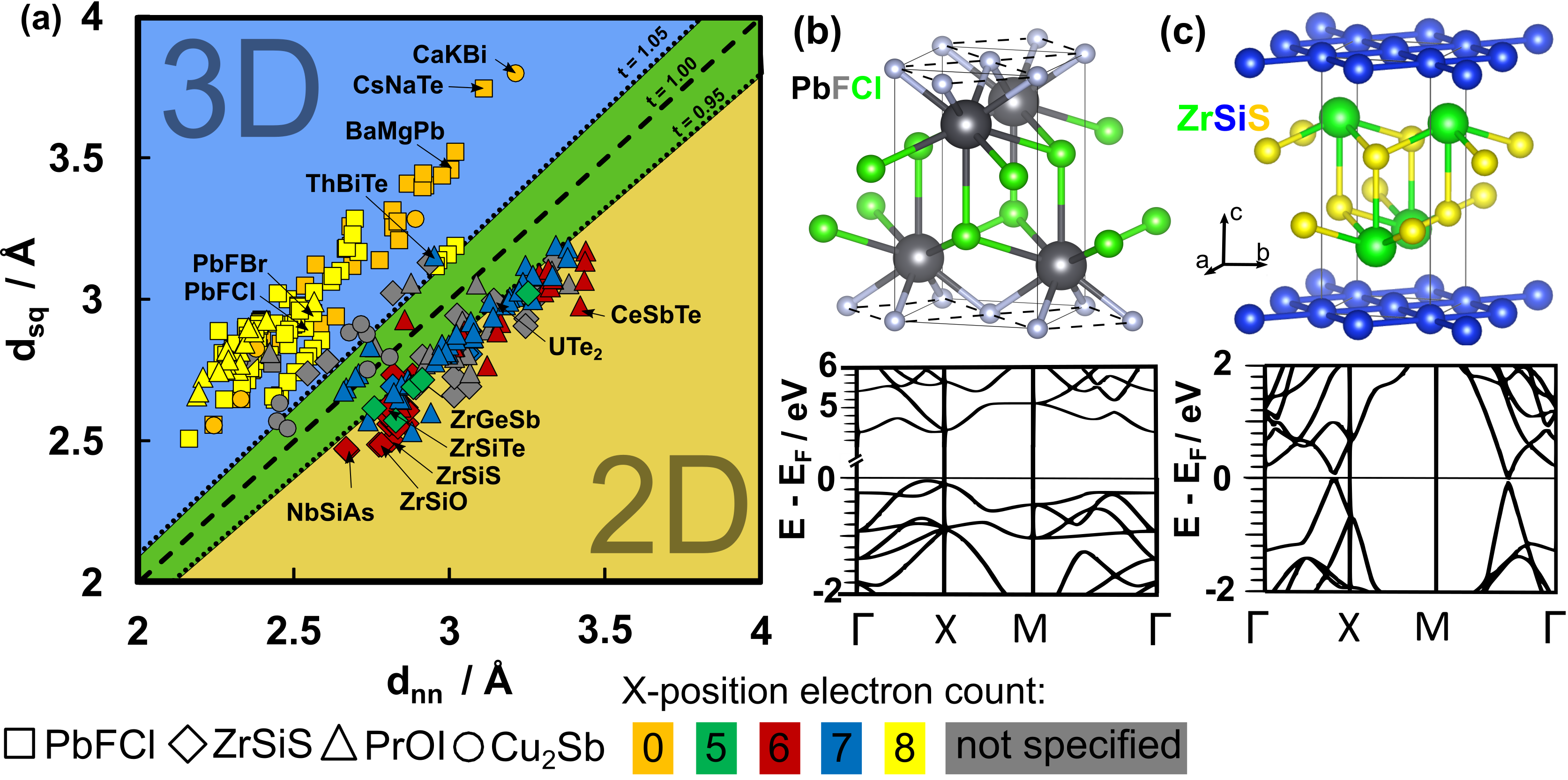}
  \caption{Graphical representation (a) for the \textit{t}-values of compounds categorized in \ce{PbFCl}, \ce{ZrSiS}, \ce{PrOI} or \ce{Cu2Sb} structure types by the ICSD. The color code represents the electron count of the MG square-net atoms. In (b) and (c) the crystal structure and band structure for \ce{PbFCl} and \ce{ZrSiS} are shown, respectively.}
  \label{MXZ-MG_data}
\end{figure}

Figure \ref{MXZ-MG_data} shows a graphical representation of the tolerance factor for MXZ-type (i.e 1:1:1 composition) square-net materials in \textit{P}4/\textit{nmm}. The structural information was taken from the \textit{Inorganic Crystal Structure Database} (ICSD);\cite{ICSD} more information about our data handling can be found in the SI. The compounds are grouped into similar structure types, according to their description in the ICSD: PbFCl, ZrSiS (including ZrSiTe, AmTe$_{2-x}$ and UP$_2$), PrOI, and Cu$_2$Sb.
The plot is divided into three areas by dashed lines, which represent \textit{t} = 1, \textit{t} = 1.05, and \textit{t} = 0.95. In the blue area (\textit{t} $>$ 1.05), the 4$^4$-net is not isolated and the structure should be considered 3D.  In the yellow area (\textit{t} $<$ 0.95), the $4^4$-net is isolated and the structure has 2D character. The green area shows the region where the geometric analysis remains ambiguous because it represents the boundary between isolated and strongly bound $4^4$-nets.

In the case of MXZ-type phases in \textit{P}4/\textit{nmm}, the compounds cluster along two lines:
Around \textit{t} = 1.2 and just under \textit{t} = 0.95. Based on electron count, the compounds with higher \textit{t} are almost exclusively ionic and electron-precise; a few gray dots in the plot indicate compounds where electron counting is ambiguous. In agreement with the tolerance factor the electronic structures of these compounds are either listed as semiconductors/insulators or electron-precise semimetals in \textit{Materials Project}\cite{MatProj,Square_Review}.
Meanwhile, in the 2D area of the plot all materials are electron deficient - the electron count per $4^4$-net atom ranges between 5 and 7, suggesting that chemical bonding plays a role. Details about how electrons were counted can be found in the SI. 

While six electrons per atom in the net is the ideal electron count for hypervalent bonds, there is some degree of flexibility. \cite{a2000hypervalent, MarrAgandZn} Of course, a deviation from 6 electrons will move the Fermi level away from the band crossings points, but the band structure remains inverted. An electron count of seven should destabilize the square-net and indeed most data points, marked blue in Figure \ref{MXZ-MG_data}, are either solid solutions with electron counts slightly above 6.5 or lanthanide dichalcogenides such as LaTe$_2$, which are known to exhibit charge density waves (CDWs).\cite{LaTe2_CDW,kikuchi1998electronic,jung2000tunneling,dimasi1996stability,kang2006charge,kang2012fermi,shim2004electronic, lee2015angle} Thus, those compounds, although listed as square-net materials in the ICSD, in reality crystallize most often in distorted structures. Recently, the evolution of these distortions in GdSb$_x$Te$_{2-x-\delta}$ with respect to $x$ was discussed in the context of TSM properties.\cite{lei2019charge} 
In addition to these compounds with seven electrons on the square-net, two compounds with five electrons are known, namely ZrGeSb (\textit{t} = 0.93) and TiGeSb (\textit{t} = 0.95). The rest of the green data points represent solid-solutions with electron counts between 5.0 and 5.5 per square-net atom.
Nonetheless, all compounds for which band structures were available in databases\cite{vergniory2019TMD} within the yellow area exhibit an electronic structure that resembles the TB model of the square-net.
We can deduce that for MXZ phases in \textit{P}4/\textit{nmm}, the tolerance factor separates materials with potential for hypervalent bonds from electron-precise compounds. The compounds in the yellow region are mostly established TSMs, for example ZrSiS and related compounds with M=Zr or Hf, X=Si, Ge, or Sn, and Z=O, S, Se, or Te.\cite{ZrSiSschoop2016,hu2016evidence,takane2016dirac,hu2017quantum,lou2016emergence,topp2016non} Other noteworthy compounds in the region are the \textit{Ln}SbTe family (\textit{Ln} = lanthanide).\cite{schoop2018tunable,hosen2018discovery,singha2016low,weiland2019band} On the zone border lies the topological superconductor \ch{UTe2}.\cite{ran2019nearly,aoki2019unconventional,miyake2019metamagnetic}  The orthorhombic phase of \ch{UTe2} was studied intensively last year because it is believed to be a strongly correlated triplet paired superconductor and could be the most promising example for topological superconductivity, which is a necessary requirement for the construction of topological qubits. The tetragonal phase has a \textit{t} = 0.95, while the orthorhombic phase contains linear Te chains, which fulfill Papoian's and Hoffmann's requirements for hypervalent bonds in linear chains.\cite{a2000hypervalent} This shows that the tolerance factor can also identify correlated topological materials.
Thus, MXZ-type phases in \textit{P}4/\textit{nmm} serve as a benchmark to show a link between hypervalency and inverted band structures.
The grey data points, which indicate compounds with ambiguous electron count, can be classified by their position in the map -- compounds with high \textit{t} can be considered electron-precise. For example, the whole solid-solution series of (Sr,Ba)F(Cl,Br), which are expected to be ionic compounds, has a \textit{t} $>$ 1.\cite{hagemann1993crystallochemical} 

While in MXZ phases in \textit{P}4/\textit{nmm} the tolerance factor did not identify new TSMs, it clearly separated those agreeing with the square-net TB model from electron-precise compounds. However, the ZrSiS-type compounds are a popular system and most known phases were investigated over the last several years,\cite{Square_Review} creating a set of phases with known topological behavior. These phases can be considered as a ``calibration'' data set. We now proceed to apply the $t$ factor to other phases that contain square-nets to use it as a predictive tool.
 
\section{Predicting TSMs Using the Tolerance Factor}
\subsection{MX$_1$X$_2$Z Compounds in \textit{P}4/\textit{nmm}}

The \textit{P}4/\textit{nmm} space group additionally contains compounds with MX$_1$X$_2$Z composition. Examples are the structure types ZrCuSiAs and \ce{HfCuSi2}.

\begin{figure}
  \includegraphics[width=0.80\textwidth]{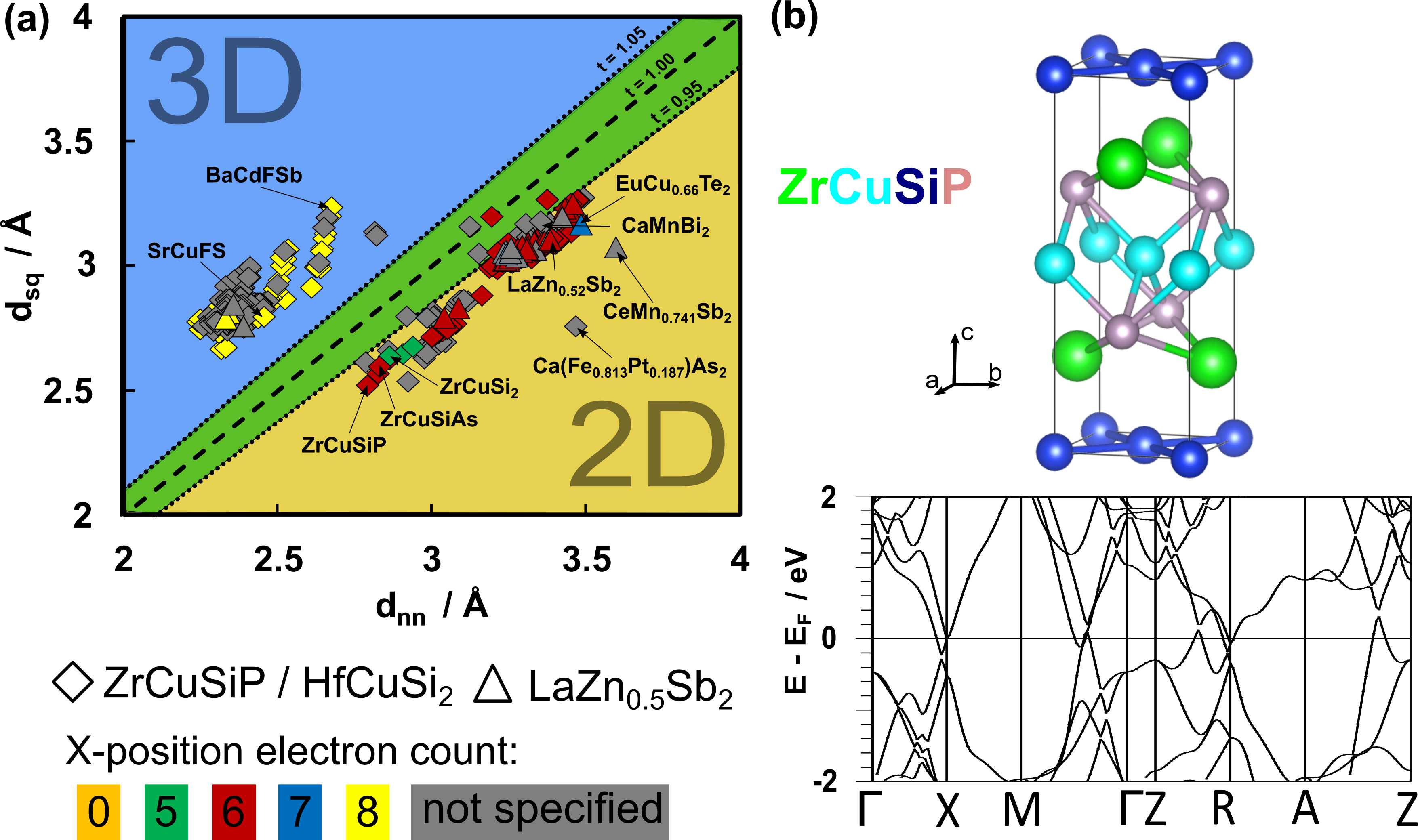}
  \caption{Graphical representation (a) for the \textit{t}-values of compounds categorized in ZrCuSiAs-\ce{HfCuSi2} or \ce{LaZn$_{0.5}$Sb2} structure types by the ICSD. The color code represents the electron count of the MG square-net atoms. In (b) the crystal structure and band structure for ZrCuSiP is shown.}
  \label{MXXZ_MG-data}
\end{figure}

Several materials in this structure type have been reported to be TSMs, these are: \ch{CaMnBi2}\cite{CaMnBi2wang2012}, \ch{YbMnPn2},\cite{YbMnBi2borisenko2015time, YbMnSb2kealhofer2018observation} \ch{LaCuSb2},\cite{LaCuSb2} and \ch{LaAgPn2}\cite{LaAgSb2myers1999haas,LaAgBi2wang2013quasi} with (Pn = Sb, Bi). Additionally, there is  an intriguing off-stoichiometric structure type: LaZn$_{0.5}$Sb$_2$. All these structure types have in common that they have two different $4^4$-nets. One is composed of a main group element (X$_2$); only this one is considered in our analysis. Figure \ref{MXXZ_MG-data} shows a graphical representation of the tolerance factor for these structure types. Similarly, electron-precise compounds have a tolerance factor larger than 1, whereas electron-deficient compounds have a tolerance factor smaller than 1. 
Out of the 379 chemically unique compounds in the ZrCuSiAs-\ch{HfCuSi$_2$} and \ch{LaZn$_{0.5}$Sb2} structure types, 165 are potentially  TSMs (\textit{t} $\leq$ 0.95). This pool of candidates includes all previously reported TSMs in the \ce{HfCuSi2} structure type such as \ce{CaMnBi2} ($t=0.95$). 
Electron counting reveals that most of these compounds have around six electrons per square-net atom. Some exceptions are compounds where a tetrel forms the $4^4$-net, such as \ce{ZrCuSi2} (five electrons per net atom) or EuCu\textsubscript{0.66}Te\textsubscript{2} (formally 6.66 electrons per net atom, though marked as 7 in the plot).

In the following we will highlight some potential new TSMs. ZrCuTtPn (Tt = tetrel, Pn = pnictide) are quaternary compounds, which meet the criteria for hypervalent bonding and have not yet been discussed with respect to topology to the best of our knowledge.

ZrCuSiP has a tolerance factor of 0.90 and ZrCuSiAs one of 0.92. Both compounds contain a Si $4^4$-net with six electrons per Si atom. 
The calculated band structures (Figure \ref{MXXZ_MG-data} (b)and S1) of both compounds display linear band crossings along $\Gamma$X and $\Gamma$M (and at the respective momenta in the $k_z=\pi$ plane); in general the band structure is closely related to that of ZrSiS.
For comparison, we also show in Figure S1 the band structure of \ce{ZrCuSi2} ($t$ = 0.92), with five electrons per $4^4$-net atom, and SrCuFS ($t$ = 1.14), which is electron-precise. While \ce{ZrCuSi2} contains some linear band crossings, the electronic structure is convoluted with several trivial bands, which can be explained by its electron count. SrCuFS is a semiconductor, as expected. Combined with electron counting the tolerance factor makes predictions of TSMs with ``clean" band structures possible, meaning that no or few other bands cross the Fermi level besides the linear band crossing. While the chemical logic explains why ZrCuSiAs and ZrCuSiP are TSMs, they were also detected by earlier high-throughput DFT calculations and are indeed listed in the various available "catalogues" of topological materials .\cite{vergniory2019TMD,materiae,tang2019comprehensive}

\begin{figure}
  \includegraphics[width=0.80\textwidth]{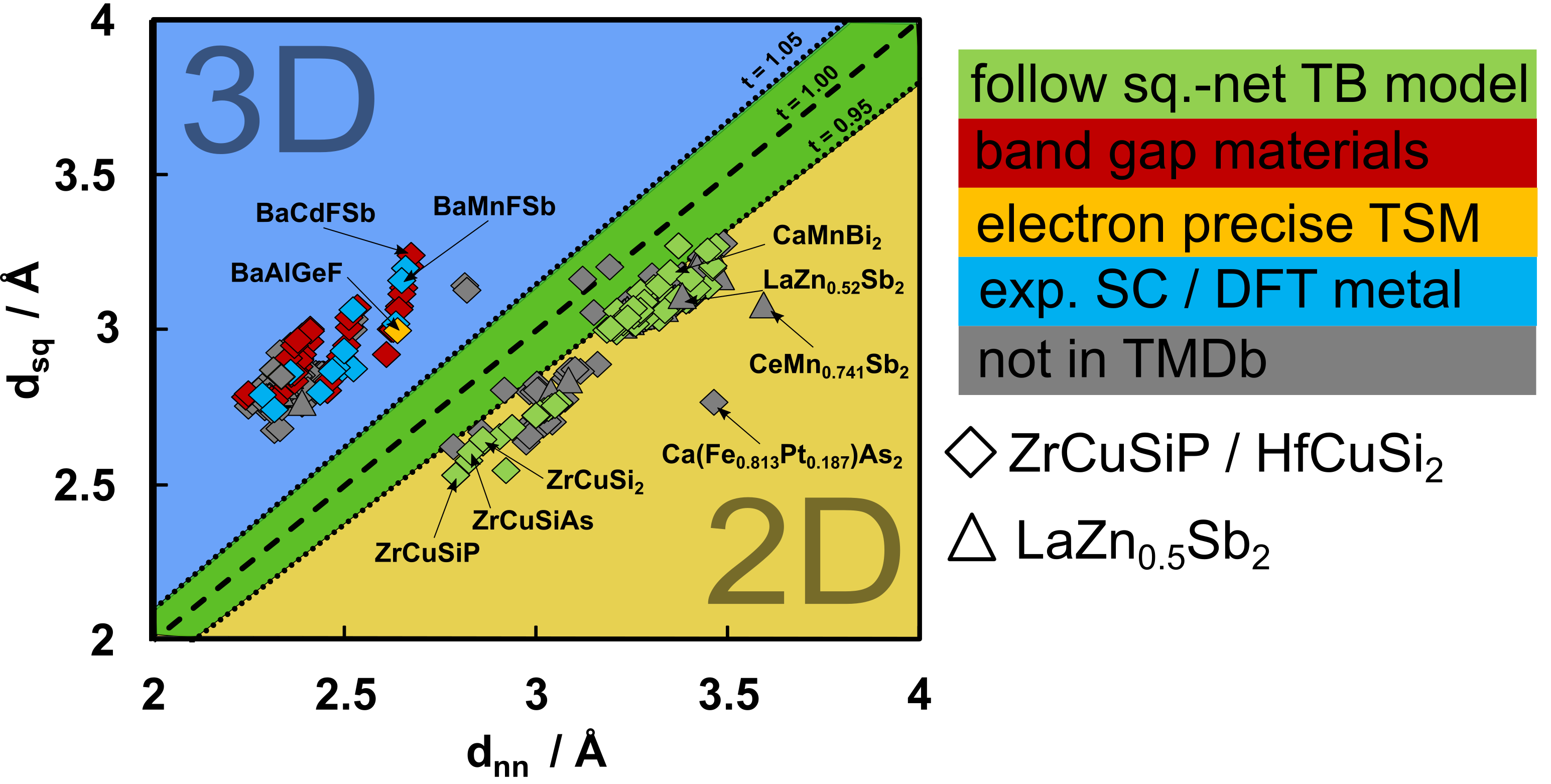}
  \caption{Graphical representation for the \textit{t}-values of compounds categorized in ZrCuSiAs-\ce{HfCuSi2} or \ce{LaZn$_{0.5}$Sb2} structure types by the ICSD. The color code represents the electronic classification of these compounds by TMDb.}
  \label{brute}
\end{figure}

At this point, we want to make a comparison to these catalogues; specifically the \textit{Topological Materials Database} (TMDb).\cite{vergniory2019TMD,bradlyn2017topological} As mentioned before, the band structures found in these databases are based on standardized DFT calculations and use structural data from the ICSD. We chose the ZrCuSiAs-HfCuSi$_2$ structure type to compare the results of using the tolerance factor to the information found in the database to determine the factor's merit. The results are summarized in Figure \ref{brute}. It contains the same data as Figure \ref{MXXZ_MG-data}, but the colors refer now to the electronic structure of an individual material (as given in the TMDb), instead of the electron count.  We distinguish between five different characteristics in the electronic structure: Green data points indicate materials for which the GGA-based DFT calculation provided in the TMDb results in a band structure that exhibits features similar to the TB model for the square-net. All green points have a tolerance factor smaller than one. Red data points indicate compounds where GGA-based DFT predicts a band gap. Orange data points indicate compounds that we classify as electron-precise semimetals, where, in the calculated band structures, band crossings appear accidentally between empty and filled shells, and where no experimental reports on electronic properties exist. Note that for these compounds higher-level DFT with hybrid functionals could open a band gap. Blue data points indicate compounds that are electron-precise semimetals within DFT, but these compounds have been experimentally reported to have a band gap. One such example is the BaMnFPn family of compounds.\cite{plokhikh2019layered} All red, blue, and orange points have tolerance factors larger than one.
Finally, grey data points indicate materials that are not listed in the TMDb. 

Grey data points with low tolerance factors will be worth studying for their potential to be TSMs.
 These compounds are of interest for chemists, because they exhibit mixed occupancies or statistical vacancies in the crystal structure, causing their omission in high-throughput searches. However, material design is often driven by doping or the synthesis of solid-solutions.
This is where the heuristic approach shines, since it allows for categorization of compounds by crystal structure alone, and there is no additional computational overhead associated with considering disordered structures.

One interesting example is the \ce{LaZn$_{0.5}$Sb2} structure type, which exhibits statistical vacancies. 

Most of these compounds follow the formula \ce{LnTM$_{1-x}$Pn2} with a partial occupancy of the X$_1$ position. The electron count in these compounds is stabilized by the vacancies to yield approximately six electrons per $4^4$-net Sb atom. These compounds highlight the large driving force for forming hypervalent bonds with six electrons. TMs in this structure type can exist in highly unusual oxidation states; one example being Mn$^+$ in CeMn$_{0.9}$Sb$_2$.\cite{a2000hypervalent} 

These compounds fulfill the requirements for hypervalent bonds and should be TSMs. Due to non-stoichiometry, verifying the prediction with DFT-based methods is more challenging, which is why this class of compounds does not appear in topological materials databases. Yet, fully occupied variants of this structure type, such as \ce{Ln(Cu$^+$/Ag$^+$)Pn2}, have been reported to be TSMs.\cite{LaAgBi2wang2013quasi, RAgSb2_dHvA, LaAgSb2myers1999haas, LaTeSb2_dirac, LaAgBi2_dirac, CeTBi2_dAvH, CeAgSb2_SvH, LaCuSb2}

\begin{figure}
  \includegraphics[width=0.80\textwidth]{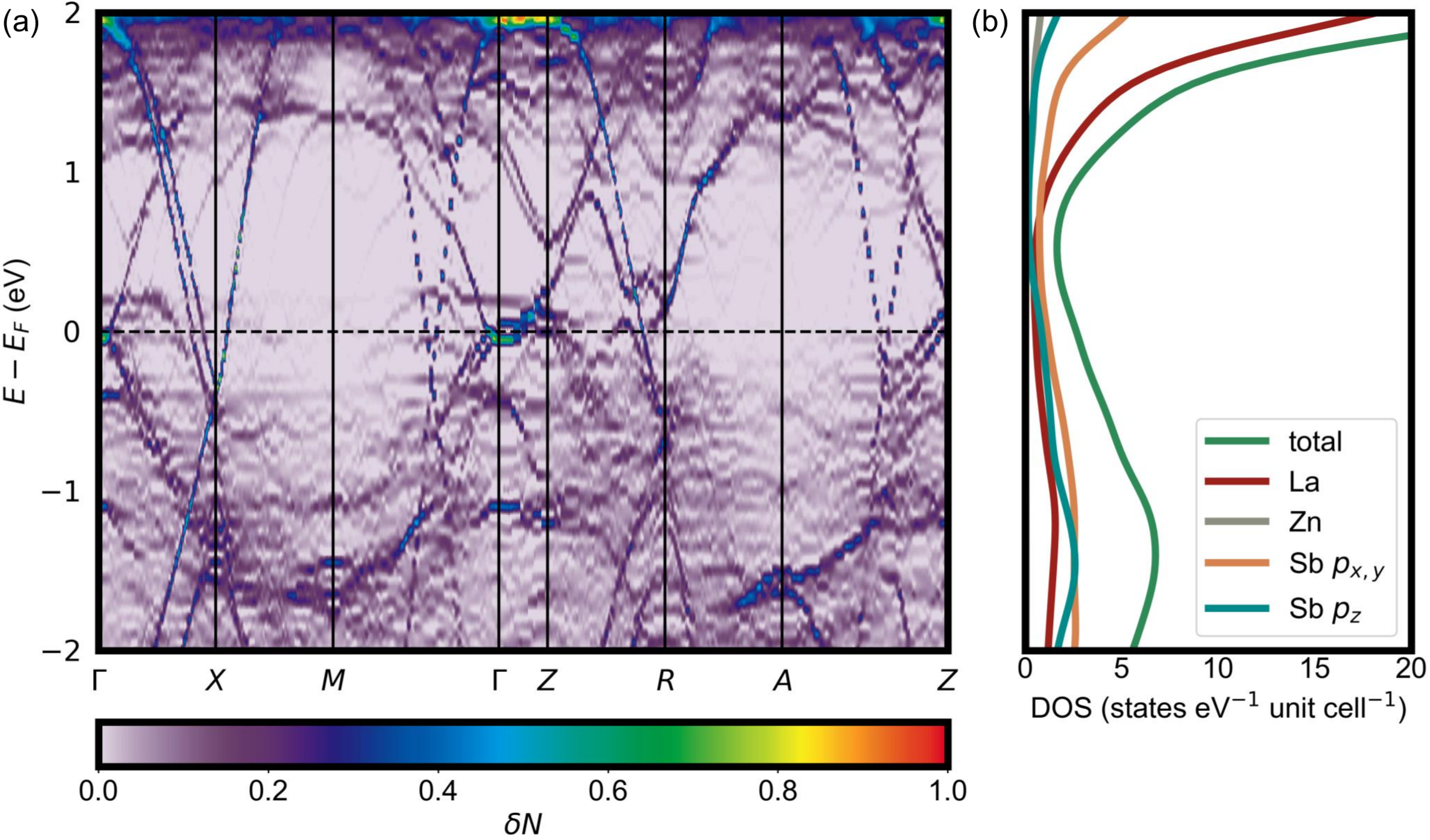}
  \caption{Electronic structure from a 112-atom supercell calculation for \ce{LaZn$_{0.5}$Sb2}: (a) unfolded electronic band structure, the color map represents the band density, $\delta$N. (b) Density of states.}
  \label{LaZn05Sb2_EBS_and_DOS_112_atom}
\end{figure}

In order to demonstrate the predictive value of the the tolerance factor in non-stoichiometric materials, we have performed such a calculation for 56-atom and 112-atom supercells of \ce{LaZn$_{0.5}$Sb2} in structures with representative correlations (special quasi-random cells). Calculation details are provided in the supporting material. Since the supercells are much larger in real space than the \ce{LaZn$_{0.5}$Sb2} unit cell, the corresponding supercell Brillouin zone is contracted in reciprocal space. The electronic bands from the supercell calculation must be ``un-folded'' into the primitive cell Brillouin zone to recreate the band structure. Figure \ref{LaZn05Sb2_EBS_and_DOS_112_atom} (a) presents the primitive electronic band structure of the 112-atom supercell (note that this unit cell still contains two square-net atoms and still has a folded band-structure if compared to a single atom unit cell).  The color map represents the band density, $\delta$N. Brighter bands with $\delta$N $>$ 0.5 mainly correspond to a primitive cell with close to six electrons per Sb and Fermi level near the Dirac cones as seen by comparing to the band structures shown in the SI (e.g., ZrCuSiAs, Figure S1). Disorder is reflected in both band smearing and the large number of bands with small $\delta$N that fill in many of the gaps in the original band structure. These small $\delta$N bands correspond to local bonding states in the supercell, which do not result from the primitive cell structure, and therefore do not map back onto the primitive cell bands. The electronic structure near the Fermi level band crossings is largely unaffected by the disorder in the supercell. A simple explanation for this robustness is provided by examining the orbital-projected density of states in Figure \ref{LaZn05Sb2_EBS_and_DOS_112_atom} (b). As expected, TSM bands near the Fermi level are mainly generated by Sb states, especially Sb $p_x$,$p_y$-orbitals. Since the square-net is preserved in the disordered structure, the Sb-Sb bonding is largely unaffected by disorder. The disordered states below the Fermi level, including the bands with greatest broadening, can be seen to originate from a combination of La and Sb states. Sb $p_z$-states, which point out of the MG square-net plane and into the disordered region, play a larger role here. The breakup of the Zn square-net does not strongly affect the near-Fermi level electronic structure because none of the electronic bands near $\mathrm{E_F}$ derive primarily from Zn orbital states; the bulk of the Zn $d$ band lies about 7\,eV below the Fermi level. Overall, this calculation confirms the value of tolerance factor analysis as a disorder-independent electronic structure prediction tool. For \ce{LaZn$_{0.5}$Sb2} and other disordered compounds in which the disorder appears outside of the main-group bonding $4^4$-net, the effect of doping on the disordered site can be reasonably approximated by considering changes in the electron count and square-net bonding. This is reflected in the experimental atomic site distances, which adjust the electron filling and band dispersion of the near-Fermi level electronic structure. We also verified the metallic nature of \ce{LaZn$_{0.5}$Sb2} by growing crystals and measuring their resistance (see SI).

\subsection{More Complex Square-Net Structures}

So far we have only discussed materials in space group \textit{P4/nmm}. However, there are many more structure types in different space groups that feature $4^4$-nets. A list of all structure types that we investigated, as well as plots of their tolerance factors are given in the SI. The list should not be seen as complete, but as a representative collection of $4^4$-nets in different space groups. This final section highlights some findings in our analysis of these additional structures.

Recently, it was shown that \ce{GdTe3} (\ce{NdTe3} structure type in space group \textit{Bmmb}) exhibits an extraordinarily high electron mobility \cite{lei2019high}. The high mobility observed in these crystals is related to the tellurium based $4^4$-net in GdTe$_3$. Its tolerance factor is \textit{t} $=$ 0.93 and isostructural compounds show similar values, ranging between \textit{t} $=$ 0.89 and \textit{t} $=$ 0.96. In this structure type, the Te square-net has a formal electron count of 6.5 electrons/Te, which has two effects. First, the Fermi level is above the crossing point, but still in the region of linearly dispersed bands. Second, the higher electron count causes the structure to distort; NdTe$_3$-type materials have been known to undergo incommensurate CDWs for some time \cite{dimasi1995chemical,ru2008haas,malliakas2006divergence,malliakas2005square}. Nonetheless, the CDW does not result in a semiconducting state and steep bands remain after the distortion.\cite{lei2019high} Therefore, the tolerance factor can be used to identify materials with exciting electronic properties, even if a structural distortion exists. Our analysis indicates the similar \ce{Nd2Te5} structure type (\textit{t} values between 0.90 and 0.91) to be a logical next step for future studies in lanthanide telluride systems.

Another noteworthy class of compounds are materials with the idealized composition \ce{Ln2TM4Pn5} in the \ce{La2Fe4Sb5} structure type\cite{weiland2019role, watkins2016emergence,phelan2012discovery}
(\textit{t} between 0.91 and 0.93; 11 compounds). These compounds are known to exhibit vacancies in the TM and Pn positions, including flux atoms and itinerant atoms, which has prevented their inclusion in structure catalogues. The unit cell contains four TM square-nets (Fe or Fe/Co) and two equivalent Pn square-nets (Sb or Sb/Bi). The electron count is controversial due to the oxidation state of Fe being ambiguous, but \textit{t}-value analysis, which does not require exact knowledge of vacancy-ordering or exact oxidation states, indicates that these materials are promising TSM candidates.

On a final note, \ch{CaBe2Ge2} structure type shows limitation of the tolerance factor. The band structure of \ch{CaBe2Ge2} shows the familiar features of a square-net material, however, the smallest \textit{t} value for this structure type is 1.08 (\ch{NdLi2Sb2}). A closer look reveals an electronic interaction between the two square-nets in the unit cell and their surrounding atoms (see SI).\cite{zheng1986donor} Thus, a different network, which is composed of edge sharing tetrahedra, has to be considered in this case. According to Hoffmann, this motif can also feature linearly dispersed band crossings \cite{zheng1986donor}. Therefore, the ``false negative'' result does not reduce the predictive power of the tolerance factor, but shows the need for further development of additional tolerance factors for different bonding geometries, which will be addressed in future work.

\section{Conclusion}
We have derived a tolerance factor that can predict topological materials in square-net materials. The tolerance factor is based on assumptions about chemical bonding, strongly implying the importance of chemical heuristics for the field of topological matter.
Applying the tolerance factor, we present topological semimetal candidates in a variety of structure types. These predictions are solely based on crystallographic information, avoiding the limitations of high-throughput DFT calculations. Therefore compounds with statistical vacancies, such as LaZn$_{0.5}$Sb$_2$, can be discovered to be topological semimetals with this method. Correlations such as magnetic order should also not affect the tolerance factor, suggesting that chemical methods can be a strong tool for moving the field of topological matter beyond the single particle limit.

\begin{acknowledgement}

This research was supported by the Arnold and Mabel Beckman Foundation through a Beckman Young Investigator grant awarded to LMS. The authors acknowledge the use of Princeton's Imaging and Analysis Center, which is partially supported by the Princeton Center for Complex Materials, a National Science Foundation (NSF)-MRSEC program (DMR-1420541).
We acknowledge use of the shared computing facilities of the Center for Scientific Computing at UC Santa Barbara, supported by NSF CNS-1725797 and the NSF MRSEC at UC Santa Barbara, NSF DMR-1720256. The UC Santa Barbara MRSEC is a member of the Materials Research Facilities Network (www.mrfn.org). SMLT has been supported by the National Science Foundation Graduate Research Fellowship Program under Grant No. DGE-1650114. Any opinions, findings, and conclusions or recommendations expressed in this material are those of the authors and do not necessarily reflect the views of the National Science Foundation.
AT was supported by the DFG, proposal No. SCHO 1730/1-1.

\end{acknowledgement}

\begin{suppinfo}

Experimental procedures, electron counting rules and
characterization data for all new compounds as well as a list of all analyzed compounds.

\end{suppinfo}

\bibliography{Klemenz_2020}

\end{document}